# Optical Separation of Mechanical Strain from Charge Doping in Graphene

Ji Eun Lee,[†] Gwanghyun Ahn, Jihye Shim, Young Sik Lee and Sunmin Ryu*

Department of Applied Chemistry, Kyung Hee University, Yongin, Gyeonggi 446-701, Korea

[†]Current address: Korea Research Institute of Standards and Science, Daejeon 305-340, Korea

*E-mail: sunryu@khu.ac.kr

**Abstract**

**Graphene, due to its superior stretchability, exhibits rich structural deformation behaviors and its strain-engineering has proven useful in modifying its electronic and magnetic properties. Despite the strain-sensitivity of the Raman G and 2D modes, the optical characterization of the native strain in graphene on silica substrates has been hampered by excess charges interfering with both modes. Here we show that the effects of strain and charges can be optically separated from each other by correlation analysis of the two modes, enabling simple quantification of both. Graphene with in-plane strain randomly occurring between -0.2% and 0.4% undergoes modest compression (-0.3%) and significant hole doping upon thermal treatments. This study suggests that substrate-mediated mechanical strain is a ubiquitous phenomenon in two-dimensional materials. The proposed analysis will be of great use in characterizing graphene-based materials and devices.**

**Introduction**

Since the first isolation of graphene from graphite,[1] atomically thin membranes of various materials have been studied revealing novel electronic,[1,2] optical,[2,3] chemical,[4-6] and mechanical[7] properties distinct from those of their bulk counterparts. Many of the findings led to proposals of unique applications, for graphene in particular, such as nanoelectronics,[8] transparent conductive electrodes,[9] high-performance composites,[10] etc. Each of the applications requires controllable modification of the material properties of graphene, for instance, electronic band gap,[8] sheet resistance,[9] or dispersibility.[11] Mechanical strain ($\varepsilon$) has also been predicted useful in implementing energy gaps in graphene under triangular stress[12] and confinement effects[13] like one-dimensional channels without physical cutting inducing unwanted charge localization on edges.[14] More recently, it was shown that non-uniform strain generates pseudo-magnetic field of 300 T pointing to a new application.[15]

On the other hand, strain can be induced uncontrollably during various processes involved in preparation of graphene sheets and devices. In particular, thermal treatments tend to generate in-plane strain due to difference in the thermal expansion coefficients (TEC) of graphene and underlying substrates. For example, graphene epitaxially-grown on SiC[16] above 1100 °C exhibits substrate-induced compressive strain ($\varepsilon \sim -1\%$) at room temperature.[17] Upon annealing at 300 °C, graphene on $SiO_2$ substrates undergoes drastic structural deformation forming sub-nm high ripples with a lateral quasi-period of several nm,[4] implying the presence of corrugation-induced strain.[18] A similar thermal rippling was observed on a larger length scale in graphene suspended over a trench.[19] Even pristine graphene on $SiO_2$ substrates, prepared by mechanical exfoliation of graphite, is deformed[4] on the nm length scale due to the *ultrastrong* adhesion[20] with the undulating substrates. Although many scanning tunneling microscopy studies have revealed corrugation[18,21-23] and related strain,[18,23] however, quantitative characterization of the native strain and its behavior upon thermal stress has been rare.[4,24]

Raman spectroscopy has been a useful tool in characterizing strain in crystalline and semi-crystalline materials[25] since changes in lattice constants lead to variations in phonon frequencies. The strain-sensitivity of the Raman frequencies of G ($\omega_G$) and 2D ($\omega_{2D}$) modes have been determined for graphene under uniaxial or biaxial stress by several groups with the resulting Gruneisen parameters in agreement with the theoretical predictions.[26-31] Both $\omega_G$ and $\omega_{2D}$ are also strongly dependent on the extra charges induced by either electrical[32,33] or chemical methods[4,6,34] due to the static effects on the bond lengths and non-adiabatic electron-phonon coupling.[35] The bimodal sensitivity of $\omega_G$ and $\omega_{2D}$ complicates independent determination of either of strain or charge density ($n$), which typically requires prior knowledge of the other. While the wide distribution in $\omega_G$ of mechanically exfoliated graphene samples were attributed to charge impurities,[36] for instance, it is not known how much native strain contributes to the variation of $\omega_G$.

Here we demonstrate that the concurrent native strain and charge doping in graphene can be determined separately from each other by Raman spectroscopy. Extensive 2-dimensional Raman analysis shows that most of pristine graphene sheets exhibit strain in the range of -0.2% ~ 0.4%, which varies gradually on the length scale of several microns. The native strain is relieved and becomes compressive when annealed at 100 °C or above, showing rich thermal transformation behaviors.

**Results**

**Strain- and charge-sensitivity of $\omega_G$ and $\omega_{2D}$.** First, we show that the pristine samples exhibit spatial inhomogeneity in $\omega_G$ ranging from 1573 cm$^{-1}$ to 1586 cm$^{-1}$. Figure 1a & 1b shows optical micrographs of two graphene samples deposited on SiO$_2$/Si substrates. As shown in Fig. 1c, the Raman spectra of these samples contain the two prominent peaks typifying single layer graphene, G and 2D peaks originating from the doubly degenerate zone-center phonon E$_{2g}$ mode and overtone of the TO phonons near K points in the Brillouin zone, respectively.[37] The absence of the disorder-related D peak near 1350 cm$^{-1}$ and symmetric Lorentzian line shape of the 2D peak with a width of ~25 cm$^{-1}$ are the most salient features of defect-free single layer graphene.[37] To investigate the spatial variations of the spectral features, hundreds of spectra were obtained per sample by raster-scanning the laser spot within the dashed boxes in Fig. 1a & 1b. The $\omega_G$-Raman maps shown in Fig. 1d & 1e exhibit significant random variations. While the edge regions in Fig. 1d have $\omega_G$ close to that of graphite (1581.5 ± 0.3 cm$^{-1}$; see Methods) with ~4 cm$^{-1}$ downshift in the central region, Fig. 1e exhibits an even wider distribution over an area of 35 x 15 μm$^2$. Besides the pixel-to-pixel variations, there are long-range undulations in $\omega_G$ occurring on a length scale of several μm. We attribute most of this frequency modulation to strain as will be shown below. In addition, we found that the density of native charges in strain-dominant pristine graphene is very low (≤ 4x10$^{11}$ cm$^{-2}$) when judged from various Raman spectroscopic features. (See Methods.)

To investigate the effects of thermal perturbation on the native strain, the samples in Fig. 1 were annealed at 400 $^o$C in vacuum for two hours. The Raman spectrum obtained following the annealing reveals ~25 cm$^{-1}$ increases in $\omega_G$ and $\omega_{2D}$ (Fig. 1c), which occurred throughout the whole graphene sheets as shown by the $\omega_G$–Raman maps in Fig. 1f & 1g. The annealing-induced stiffening of both Raman modes, first reported by Li et al.,[6] was attributed to hole doping caused by O$_2$ in the presence of water[4], although exact doping mechanisms still remain unclear.[4,38-42] The intensity decrease and line broadening in 2D mode of the annealed graphene (Fig. 1c) are also mainly attributed to the hole doping.[43,44] While Fig. 1f confirms the upshift in $\omega_G$ across the entire area, it also reveals that the spatial variation of $\omega_G$ has been removed upon the annealing. On the contrary, the $\omega_G$-undulation in Fig. 1e remains almost unaffected by the thermal treatment despite the annealing-induced upshift (Fig. 1g).

The puzzling spectral variations above are now presented in a different perspective in Fig. 2 to show how the pixel-to-pixel variations in $\omega_G$ correlate with those in $\omega_{2D}$. Interestingly, hundreds of data points from a given pristine sample, each corresponding to a spectral average over ~1 μm$^2$, form a linear line. Remarkably, the data sets from 8 pristine samples including three in Fig. 2 turned out to fall on a single line with a slope ($\Delta\omega_{2D}/\Delta\omega_G$) of 2.2 ± 0.2 (black dashed line). To determine intrinsic frequencies of the two modes ($\omega_G^0$, $\omega_{2D}^0$) not affected by strain or excess charges, we investigated a freestanding graphene (**F1**; Supplementary Figure S1) suspended across a circular well (7 μm in diameter and 5 μm in depth). The green circle in Fig. 2 indicates the values averaged over a freestanding area of 16 μm$^2$, (1581.6 ± 0.2, 2676.9 ± 0.7). Freestanding graphene is known to be virtually charge-neutral with a residual charge density less than 2x10$^{11}$ cm$^{-2}$.[34] Despite the possibility of pre-tension in the suspended graphene,[45] we conclude that **F1** is essentially strain-free since $\omega_G^0$ agrees well with the aforementioned value of graphite and that of electrically neutralized bilayer graphene[46] within 0.5 cm$^{-1}$ corresponding to a biaxial strain less than ~0.01%.[30] Setting ($\omega_G^0$, $\omega_{2D}^0$) as the origin (**O**) of the $\omega_G$–$\omega_{2D}$ space, the points near **O** in Fig. 2 originate from graphene areas which are nearly charge- and strain-free like the freestanding graphene and the rest are mechanically strained or charge-doped.

Although $\omega_G$ and $\omega_{2D}$ are highly sensitive to both ***n*** and ***ε***, we note that their fractional variation due to ***n***, $(\Delta\omega_{2D}/\Delta\omega_G)_n$, is very different from that caused by ***ε***, $(\Delta\omega_{2D}/\Delta\omega_G)_\varepsilon$. Firstly, biaxially strained graphene, either compressive or tensile, shows a fairly large ratio of $(\Delta\omega_{2D}/\Delta\omega_G)_\varepsilon^{biaxial}$: three groups reported experimental values of 2.45,[47] 2.63,[31] and 2.8,[30] whereas theory predicted slightly smaller values of 2.25[48] and 2.48.[27] For graphene under uniaxial stress, however, $(\Delta\omega_{2D}/\Delta\omega_G)_\varepsilon^{uniaxial}$ depends on the direction of the strain with respect to the crystallographic axes of graphene. Because of the strain-induced symmetry breaking, the G (2D) mode of graphene under uniaxial stress splits into G$^-$ (2D$^-$) and G$^+$ (2D$^+$).[26,27] When graphene is strained along the zigzag (arm-chair) directions, $(\Delta\omega_{2D}^-/\Delta\omega_G^-)_\varepsilon$ = 2.05 (1.89) and $(\Delta\omega_{2D}^+/\Delta\omega_G^+)_\varepsilon$ = 2.00 (3.00).[29] When the observed G (2D) peaks are resultant of the G$^-$ (2D$^-$) and G$^+$ (2D$^+$) peaks that are not resolved because of insufficient splitting, $(\Delta\omega_{2D}/\Delta\omega_G)_\varepsilon$ for the zigzag (arm-chair) directions can be approximated as an average of $(\Delta\omega_{2D}^-/\Delta\omega_G^-)_\varepsilon$ and $(\Delta\omega_{2D}^+/\Delta\omega_G^+)_\varepsilon$, 2.02 (2.44). However, since the native strain in mechanically exfoliated graphene can be aligned along any direction between the zigzag and arm-chair axes, $(\Delta\omega_{2D}/\Delta\omega_G)_\varepsilon^{uniaxial}$ is expected to lie in the range of 2.02~2.44, which is in an excellent agreement with 2.2 ± 0.2 obtained from the eight samples. Secondly, the effects of ***n*** are dependent on the sign of the charges because of their static effects on bond-length and are more pronounced for $\omega_G$ than $\omega_{2D}$ because of the non-adiabatic electron-phonon coupling.[32,33,35] Hole doping induced by electrical gating leads to a quasi-linearity (($\Delta\omega_{2D}/\Delta\omega_G)_n^{hole}$ = 0.75 ± 0.04) between $\omega_G$ and $\omega_{2D}$ as shown by the red solid line in Fig. 2 (Supplementary Figure S2 and Supplementary Methods A), while $\Delta\omega_{2D}/\Delta\omega_G$ for electron doping becomes more nonlinear for high charge density as depicted by the blue solid line.[49] However, we exclude the possibility of electron-doping in the current studies since many charge transport[50] and Raman scattering[36] studies have observed hole doping predominantly in pristine[36] and annealed[4] graphene. Thus, based on the negligible charge density in these samples (Methods), we conclude that the linear variations of $\omega_G$ and $\omega_{2D}$ are due to native strain in graphene. While our results are in better agreement with the scenario of uniaxial strain, the presence of biaxial strain or mix of both could not be excluded viewing the disagreement in experimental $(\Delta\omega_{2D}/\Delta\omega_G)_\varepsilon^{biaxial}$ and discrepancy between experiment and theory. The presence of native strain also leads to an interesting question on "*strain coherence length*", how large the strained domains are or how far the direction of the strain is maintained, which is beyond the scope of the current studies. However, a recent STM study suggests that the length scale can be as small as a few nm for graphene on $SiO_2$ substrates,[23] which is also consistent with the spatial distribution of in-plane atomic displacements resulting from thermal fluctuation.[51]

**Vector decomposition of *ε* and *n*.** Now, it is logical to extract the contribution by *ε* or *n* for a given point in the $\omega_G$–$\omega_{2D}$ space, $\mathbf{P}(\omega_G, \omega_{2D})$, using a simple vector model as depicted in the inset of Fig. 2: $\mathbf{OP} = a\mathbf{e_T} + b\mathbf{e_H}$, where *a* and *b* are constants, $\mathbf{e_T}$ and $\mathbf{e_H}$ are unit vectors for tensile strain $((\Delta\omega_{2D}/\Delta\omega_G)_\varepsilon^{uniaxial} = 2.2 \pm 0.2)$ and hole doping effects $((\Delta\omega_{2D}/\Delta\omega_G)_n^{hole} = 0.70 \pm 0.05)$, respectively. The $\omega_G$–$\omega_{2D}$ space is now divided into four quadrants ($Q_1 \sim Q_4$) by $\mathbf{e_T}$ and $\mathbf{e_H}$. As increasing *n* (*ε*) of an intrinsic graphene, its values of ($\omega_G$, $\omega_{2D}$) will move from **O** along $\mathbf{e_H}$ ($\mathbf{e_T}$). While $Q_4$ ($Q_1$) is attributed to tensile (compressive) strain combined with hole doping, $Q_2$ and $Q_3$ are not allowed since both of electron and hole doping should lead to increase in $\omega_G$. Thus, the variations in $\omega_G$ and $\omega_{2D}$ of the pristine graphene in Fig. 2 are mostly due to strain with negligible charge doping ($b \approx 0$) and the extent of tensile strain is typically a few times larger than that of compressive one for a given sample. Few pristine graphene sheets of smaller area, however, exhibited non-negligible doping concurrent with strain as will be shown below.

While the assumption of constant $(\Delta\omega_{2D}/\Delta\omega_G)_n^{hole}$ is approximately valid on a wider frequency range, more accurate analysis of *ε* and *n* can be performed using a theoretical prediction[35] (blue solid line in Fig. 3) and the experimental data[49] for $(\Delta\omega_{2D}/\Delta\omega_G)_n^{hole}$ (red solid line in Fig. 3) as explained below. In addition, the variation of $\omega_G$ caused by the change in *n*, thus the Fermi level ($E_F$), becomes nonlinear with respect to $\Delta E_F$ at low charge density ($|n| \sim 1\times10^{12}$ cm$^{-2}$) because of the anomalous softening[35] of G phonon occurring when $|E_F| = \hbar\omega_G/2$ (Supplementary Methods B). As a refined approach over the original vector model, Fig. 3a shows a new trajectory of $\mathbf{O}(\omega_G^0, \omega_{2D}^0)$ for hole doping (blue solid line) theoretically predicted[35] by considering the phonon anomaly for *n* ranging from 0 to $2.6\times10^{12}$ cm$^{-2}$. With increasing *n*, $\mathbf{O}(\omega_G^0, \omega_{2D}^0)$ first moves into the blue-shaded region, a forbidden area in $Q_2$, and returns back into the doping-affected area ($Q_1$) represented by the yellow shade at *n* $\sim 1.4\times10^{12}$ cm$^{-2}$. As further increasing *n*, the refined trajectory approaches $\mathbf{e_H}$ very closely as can be seen in Fig. 3b. In addition, the area of the blue region due to the anomalous softening ($\delta\omega_G^{Anom}$ in Fig. 3a) is very small compared to the yellow region on a larger frequency scale as shown in Fig. 3b. We note, however, that any given point in the blue area or on the $\mathbf{e_T}$ line, $\mathbf{A}(\omega_G, \omega_{2D})$, can be attributed to either **A'** or **A"** affected by strain. Thus, unambiguous determination of *ε* and *n* cannot be made for **A** in the blue shade and the associated errors turn out to be $\delta\varepsilon \leq 0.03\%$ and $\delta n \leq 1.5 \times 10^{12}$ cm$^{-2}$. In contrast, **B**($\omega_G$, $\omega_{2D}$) in the yellow area can be unequivocally interpreted as **B'** affected by compressive strain, thus enabling more accurate determination of $\varepsilon$ and *n*. However, the experimental path (red solid line in Fig. 3a) does not enter the blue area because of the absence of the anomalous softening of G mode as explained in Supplementary Figure S3. Over the wide range of *n* (0 ~ $1.6 \times 10^{13}$ cm$^{-2}$) shown in Fig. 3b, however, the experimental (red solid line) and theoretical (blue solid line) trajectories agree well with each other and can be well represented by the **e**$_H$ line. Thus, the refined approach in Fig. 3b enables one to disentangle the degree of strain from that of charge doping more accurately. For the pristine graphene shown in Fig. 3b, for example, it can be seen that $\varepsilon$ ranges from -0.2% to 0.4% with *n* < $1.0 \times 10^{12}$ cm$^{-2}$ for the majority of the data points. It should be noted, however, that few experimental data sets[43,49] for $(\Delta\omega_{2D}/\Delta\omega_G)_n^{hole}$ available in the literature reveal non-negligible discrepancy (Supplementary Figure S3) and thus refined experimental data will enhance the accuracy of the proposed analysis.

**Thermal modulation of $\varepsilon$ and *n*.** We demonstrate that thermal annealing in vacuum removes the native tensile strain and induces compressive strain. The most prominent change caused by the thermal annealing in Fig. 2 is that the concurrent stiffening of the G and 2D modes. The refined analysis model in Fig. 3b readily leads to quantification of strain and charge density: despite the wide distribution, most of the ($\omega_G$, $\omega_{2D}$) points of annealed **K2** lie on a line parallel to **e**$_T$ with ~80% of the thermally induced changes in $\omega_G$ found along the **e**$_H$ axis. This indicates that the O$_2$-induced hole doping activated by annealing[4] dominates the spectral changes and *n* remains relatively constant at $(1.4 \pm 0.1) \times 10^{13}$ cm$^{-2}$. The wide distributions in $\omega_G$ and $\omega_{2D}$, instead, can be attributed to compressive strain (-0.3% $\leq \varepsilon \leq$ 0), which contrasts with the fact that the native strain was mostly tensile in nature. This finding is in agreement with the annealing-induced slippage[19,24] and buckling[4] of graphene on SiO$_2$ substrates caused by the difference in TECs of both materials,[52] since the former relieves the tensile strain[24] and the latter accompanies compression.[19] We also note that thermal modulation of strain is sample-specific. For example, the spectral spreads of **K1** and **K3** in the $\omega_G$–$\omega_{2D}$ space decreased greatly upon annealing, while those of **K2** underwent only minor changes as shown in Fig. 2.

In Fig. 4 presenting the $\omega_G$-$\omega_{2D}$ graph obtained from **K4**, we varied the annealing temperature stepwise to determine the critical temperatures where the native strain starts to relax and changes into compression. It can be seen that the pristine graphene has mostly tensile strain with negligible hole doping: each **P**($\omega_G$, $\omega_{2D}$) of **K4** lies on the **e**$_T$ axis in the range between (1578, 2670) and (1582, 2678). Following the first annealing at 100 °C, most of **P**($\omega_G$, $\omega_{2D}$) moved into the range between (1580, 2676) and (1585, 2683), but still mostly along **e**$_T$. This change demonstrates that heating at 100 °C can be sufficient to remove most of the native tensile strain and compress graphene simultaneously causing slight hole doping at certain areas. While subsequent annealing at 150 °C led to more obvious hole doping in addition to further compression, repeated annealing at 200, 250, and 300 °C resulted in less significant variations in $\mathbf{P}(\omega_G, \omega_{2D})$; data for the treatments at 200 and 250 °C are not shown to avoid congestion. Further treatments at 400 and 500 °C, however, induced marked movement of $\mathbf{P}(\omega_G, \omega_{2D})$ to (1591.5 ± 1.4, 2693 ± 2.3) and (1598.4 ± 1.4, 2698 ± 2.6), respectively. The changes are largely in parallel with $e_H$, indicating emergence of strong hole doping.

We also note hysteretic effects in annealed graphene samples. Compared to the one-time annealing at 400 °C (**K1~K3** in Fig. 2), the sample (**K4**) that underwent cycles of prior annealing at lower temperatures exhibit much less changes but larger distributions in frequencies (Fig. 4) and linewidths (Supplementary Figure S4). Since the buckling of annealed graphene sheets responsible for the thermally induced hole doping[4] is dictated by adhesion and slippage of graphene on silica,[19,24] it is likely that prior history of thermal treatments affects the buckling behaviors and thus $\omega_G$ and $\omega_{2D}$. The spectral inhomogeneity increased by the repeated annealing cycles can be attributed to further structural deformation or in-situ reactions with residual gases in the vacuum system during annealing or post-annealing surface reactions[6] occurring in the ambient conditions. Since annealing is widely used in fabricating graphene transistors[50] and preparing graphene samples for various fundamental research,[18,21] the exact chemical changes made by annealing deserve careful studies in the future. The Lorentzian linewidths of G ($\Gamma_G$) and 2D ($\Gamma_{2D}$) peaks, and the 2D/G peak area ratios ($A_{2D}/A_G$) determined following each annealing cycle are also consistent with the scenario of thermally induced mechanical transformation concurrent with hole doping (Supplementary Figure S4 and Supplementary Methods C).

**Spatial mapping of $\varepsilon$ and $n$.** In Fig. 5, we demonstrate that the native strain concurrent with spatially varying charge doping can be optically mapped out using the refined vector analysis. The Raman maps obtained from the graphene sample (**K5** in Fig. 5a) reveal that $\omega_G$, $\omega_{2D}$ and $\Gamma_G$ exhibit variations of several cm$^{-1}$ across an area of 20x15 μm$^2$ respectively in Fig. 5b, 5c and 5d. Unlike the strain-dominated graphene (**K1~K4**), all $\mathbf{P}(\omega_G, \omega_{2D})$ from **K5** are scattered in $Q_4$ instead of forming a line along $e_T$ (Fig. 5g), indicating the coexistence of tensile strain and hole doping. According to the vector analysis in Fig. 2, each point of $\mathbf{P}(\omega_G, \omega_{2D})$ can then be decomposed into $\mathbf{H}(\omega_G, \omega_{2D})_{\varepsilon=0}$ and $\mathbf{T}(\omega_G, \omega_{2D})_{n=0}$, representing phonon frequencies which are not affected by $\varepsilon$ or $n$, respectively. Shown in Fig. 5e and 5f are the resulting Raman maps of $\omega_{G,\,\varepsilon=0}$ and $\omega_{G,\,n=0}$: resorting to Fig. 5g, the former reveals the spatial distribution of the hole density ranging up to 3.5x10$^{12}$ cm$^{-2}$, while the latter maps out the native strain (-0.03% < $\varepsilon$ < 0.17%). Figure 5e & 5f also reveal that the long-range distribution of the strain does not necessarily coincide with the native charge distribution. It is also to be noted that $\omega_{G, \varepsilon=0}$ and $\Gamma_G$ obey a reciprocal relation which conforms to the theoretical prediction for charge doping,[43] while $\omega_G$ and $\Gamma_G$ exhibit a much broader distribution due to the coexisting strain (Supplementary Figure S5). Similar improvement was obtained in the correlation between $\omega_{G, \varepsilon=0}$ and $A_{2D}/A_G$ (Supplementary Figure S5).

**Discussion**

The current study shows that most of graphene on silica substrates are mechanically strained and tensile strain is more frequently found than compressive one in a given sample. In this regard, graphene can be envisaged as food wrap which tends to become strained forming ripples when clinging to flat surfaces. While both tensile and compressive shear stresses can be applied to the membrane, facile buckling along the out-of-plane direction will make compressive strain less likely than tensile one. Interestingly, the extent of the native strain is the larger on average for the larger graphene flake, which may be due to the fact that larger contact area with substrates provides enhanced adhesion to resist slippage caused by in-plane stress. However, thermal perturbation as shown in the current study and varying interactions with other substrates should lead to diverse structural transformation of graphene and other newly discovered 2-dimensional materials such as $h$-BN, $MoS_2$, $MoSe_2$, *etc*. We also note that strain dominates the spectral variations over charge doping which has been considered mainly responsible for the spectral irregularities in graphene.[36] This suggests that the degree of mechanical deformation or charge doping depends on preparation methods. Insignificant chemical doping in our pristine samples could be due to different chemical and structural properties of the substrate surfaces.[38] In this regard, the presence of tensile strain in graphene may have an influence on the degree of charge doping. A recent STM study revealed that graphene is partly suspended between microscopic hills of the substrates when supported on $SiO_2$ substrates.[22] Tensile stress is likely to induce *localized suspension* of graphene which otherwise would largely conform[18] to the undulating substrates due to the van der Waals interaction.[20] (See Supplementary Methods D.) Such semi-freestanding graphene should be less sensitive to charge-doping that occurs via contact with the substrates[34,38,40] or through the corrugation-mediated mechanisms.[4,6] We also note that strain may give rise to charge inhomogeneity through rehybridization of π-σ bonds and vice versa.[53,54] However, no clear correlation was found as shown in Fig. 1 & 5 and Methods, presumably because of the insufficient spatial resolution and limited sensitivity towards charge density and strain.

The spatial distribution of native strain in mechanically exfoliated graphene samples has rarely been quantified.[23] Moreover, the strain has not been systematically considered in interpreting Raman spectra due to the competing effects of extra charges, despite the well characterized strain-sensitivities of the G and 2D bands.[26-31] Our studies demonstrate that the G and 2D Raman modes of graphene can be highly reliable in determining mechanical strain and charge density even when both coexist. The bimodal sensitivity of both modes, however, requires careful interpretation as suggested in this paper. Although many STM studies revealed structural irregularities such as buckling and strain in graphene,[18,21-23] the method is not practically useful in achieving statistical information on a length scale larger than microns. Moreover, the current studies demonstrate that graphene undergoes sample-specific hysteretic structural deformation upon thermal treatments and possibly other external perturbations. Since typical micro-fabrication[50] and STM measurements[18,21] of graphene and its devices involve various sample treatments such as annealing, transfer to substrates, polymer coating, wetting-drying, *etc.*, their effects need be considered in interpreting results. Our studies also suggest that not only graphene but also other 2-dimensional materials supported on solid substrates are generally susceptible to native strain and thermal deformation due to zero-dimension along the z-axis and different TEC's of involved materials.

Despite providing a systematic analysis, however, the current study also shows some limitations. In principle, simultaneous vector decomposition into strain, p-type, and n-type doping cannot be made unambiguously requiring that contribution of at least one component should be known or assumed. The non-zero dispersion of $\omega_{2D}$ limits the current approach to the Raman measurement obtained with 514 nm as an excitation source. Follow-up studies are being carried out with other excitation wavelengths. This work is also limited to single-layer graphene and analysis of few-layer graphene will require separate set of data. Finally, the accuracy of this approach will be directly affected by the spectral accuracy of employed spectrometers which can be tested against $\mathbf{O}(\omega_G^0, \omega_{2D}^0)$.

In conclusion, we have demonstrated that the native strain can be unambiguously determined by Raman spectroscopic analysis notwithstanding the interference from the coexisting charge doping effects. Most of the pristine graphene sheets deposited onto $SiO_2$ substrates by the mechanical exfoliation method were shown to be under in-plane stress with the resulting strain in the range of -0.2% ~ 0.4%. The native tensile strain was relieved by thermal annealing at a temperature as low as 100 °C and converted to compressive strain by annealing at higher temperatures, which also induced strong hole doping clearly resolved in the analysis. The proposed analysis should be useful for fast and reliable characterization of strain and excess charges in graphene materials and devices.

**Methods**

**Preparation and treatment of samples.** High quality graphene samples were prepared by the micro-mechanical exfoliation method[50] using kish graphite (Covalent Materials Inc.) and adhesive tape (3M). The Si substrates with 285 nm-thick $SiO_2$ layers were cleaned with piranha solutions prior to the deposition of graphene. For freestanding graphene samples, substrates with micron-scale circular wells (diameter: 2~7 μm, depth: 5 μm) were employed.[45] For thermal annealing, samples in a tube furnace evacuated to a pressure of 3 mTorr were heated to a target temperature ($T_{anneal}$) within 30 min, maintained at $T_{anneal}$ for 2 hours, and then cooled down to 23 °C for ~3 hrs.

**Raman spectroscopy.** The number of layers and crystallinity of the prepared samples were characterized by Raman spectroscopy.[37] All the Raman spectra were obtained in a back scattering geometry using a 40x objective lens (NA = 0.60) in the ambient conditions. An Ar ion laser operated at a wavelength of 514.5 nm was used as an excitation source. While the spectral width of the instrument response function was determined to be 3.0 cm$^{-1}$ from the Rayleigh scattering peak, the spectral precision and accuracy were better than 1.0 cm$^{-1}$ from repeated measurements of Raman standards. (See Methods below for detailed analysis.) For two-dimensional Raman maps, spectra were obtained every 1 μm using an x-y motorized stage. The average power of the excitation laser beam was 1.5 mW which was focused onto a spot of ~0.5 μm in diameter. No irreversible photoinduced change was detected during the measurements.

**Spectral accuracy of the Raman measurements.** Although single-grating spectrometers of Czerny-Turner type, including the one (SP2300, Princeton Instruments Inc.) employed in the current study, provide high throughput and small footprints,[55] careful calibration is required to achieve instrument-limited spectral accuracy because of the wavelength dependence of its reciprocal linear dispersion (RLD).[56] More specifically, wavelength ($\lambda$) of each CCD detector pixel needs to be expanded in series of the pixel position ($x$): $\lambda = \lambda_0 + a_1(x-x_0) + a_2(x-x_0)^2 + a_3(x-x_0)^3 + \ldots + a_n(x-x_0)^n$, where $a_n$, $\lambda_0$ and $x_0$ are constants, the center wavelength and its position on the detectors, respectively. Supplementary Figure S6a presents the wavelengths of 13 plasma lines from the Ar laser and 3 Hg atomic lines as a function of their pixel positions recorded in the CCD detector. Although the data appear well described by the linear line in blue, the first order calibration with $a_n = 0$ (n>1) leads to non-negligible error of -0.15 ~ 0.25 nm in wavelength as can be seen in Supplementary Figure S6b. We note that this amount of deviation translates into Raman shift error of 5.0 ~ -6.8 cm$^{-1}$ which is even larger than the linewidth of the Rayleigh line (3.0 cm$^{-1}$). It is also to be noted that a first order calibration using only 3 Hg lines (546.075, 576.961 & 579.067 nm) instead of the above 16 lines generates even larger error up to 0.68 nm or 18 cm$^{-1}$ across the entire detector area. When the quadratic term was included in the calibration as shown in Supplementary Figure S6b, however, the deviation remained within ±0.01 nm or ±0.3 cm$^{-1}$, which is sufficient in accuracy for the employed spectrograph with a focal length of 300 mm and a grating with 1200 grooves/mm. This suggests that extra caution needs to be paid when comparing Raman G and 2D frequencies recorded by different spectrographs with modest spectral resolving power. We further tested the accuracy by measuring the G band frequency ($\omega_G$) of thick graphite flakes: $\omega_G$ from 12 different spots out of four different samples was 1581.5 ± 0.3 cm$^{-1}$, which turned out to be within ~0.5 cm$^{-1}$ from the literature values of 1581~1582 cm$^{-1}$.[57,58] Thus, the accuracy of our measurements was conservatively claimed as 1.0 cm$^{-1}$.

**Negligible native charge density in pristine graphene.** The scheme of vector decomposition proposed in this article assumes that the variation in ($\omega_G$, $\omega_{2D}$) of the employed pristine samples except **K5** is mostly due to strain and that the contribution of charge doping is sufficiently small for graphene under tensile stress. This hypothesis is well supported by a few different spectral features of the pristine samples as shown below. Supplementary Figure S7a shows that ($\omega_G$, $\omega_{2D}$) of the three samples lies on the black-dashed line (**e$_T$**) representing graphene affected by strain, but not charge.

Moreover, the 2D/G peak area ratio ($A_{2D}/A_G$ = 5.8 ± 0.3) in Supplementary Figure S7b remains constant and very close to that ($A_{2D}/A_G^0$ = 6.2 ± 0.2) of the charge neutral freestanding sample (**F1**) while $\omega_{2D}$ varies by more than 10 cm$^{-1}$. (See Supplementary Methods E for the optical artifact caused by the substrates which affects the apparent $A_{2D}/A_G$.) Since $A_{2D}/A_G$ decreases drastically as increasing the charge density (***n***) for either type of charges,[43] Supplementary Figure S7b supports that the presented pristine graphene samples have negligible charge density regardless of the widely varying native strain.

The behavior of $\Gamma_G$ that is not affected by the optical artifact will be useful in judging the native charge density. As shown in Supplementary Figure S7c, $\Gamma_G$ of the three pristine samples lies in the range of 13.1 ± 0.7 cm$^{-1}$, which is only slightly smaller than that of the freestanding sample, $\Gamma_G^0$ = 13.9 ± 0.2 cm$^{-1}$. Since $\Gamma_G$ is sensitive to low level of charge doping due to the blockage of the non-adiabatic decay channel of the G phonon, the native charge density in the pristine samples are generally very small.[34] A quantitative estimation of ***n*** according to the model proposed by Berciaud *et al*.[34] leads to a conclusion that $\Gamma_G$ = 13.1 ± 0.7 cm$^{-1}$ translates into |***n***| < 4x10$^{11}$ cm$^{-2}$. (Note that equation 1 and Fig. S4 of the reference[34] were employed.) This level of charge density is an order of magnitude lower than the variations reported for graphene supported on SiO$_2$ substrates by others.[34,36]

The distribution of $\Gamma_{2D}$ as a function of $\omega_{2D}$ shown in Supplementary Figure S7d also supports the assumption that the pristine graphene samples of which ($\omega_G$, $\omega_{2D}$) lies on **e**$_T$ in Supplementary Figure S7a are not affected by significant level of charge doping. Das *et al.* showed that $\Gamma_{2D}$ increases by ~30% in contrast to decreasing $\Gamma_G$ when |***n***| is raised to ~2x10$^{13}$ cm$^{-2}$ by an electrical method.[43] Several groups reported that $\Gamma_{2D}$ of supported graphene samples with some level of p-type doping lies in the range of 28 ~ 30 cm$^{-1}$,[36,59] which is significantly larger than that of freestanding graphene (22.5 ~ 24 cm$^{-1}$).[34] We confirmed that $\Gamma_{2D}$ of the charge neutral freestanding graphene (**F1**) is 23.1 ± 0.2 cm$^{-1}$, and found that $\Gamma_{2D}$ of the supported samples in Supplementary Figure S7d also remains at very small values, 23.5 ± 1.2 cm$^{-1}$, indicating low level of native charge density.

**Acknowledgements**

The authors thank Duhee Yoon and Hyeonsik Cheong for sharing optical data and substrates and Taeg Yeoung Ko for helpful comments. This work was supported by the National Research Foundation of Korea (No. 2011-0031629, 2011-0027288, 2011-0010863). J.S. acknowledges the financial support from Kyung Hee University (KHU-20110209).

**Author contributions**

J.L., G.A. & J.S. performed the experiments. J.L., G.A. & S.R. analyzed the data and S.R. wrote the paper. All authors discussed the results and commented on the manuscript.

**Additional information**

**Supplementary Information** accompanies this paper at http://www.nature.com/naturecommunications.

**Competing financial interests:** The authors declare no competing financial interests.

**Figure legends**

**Figure 1. Raman maps of single layer graphene revealing large frequency variations. a**, Optical micrograph of sample **K1**. The scale bar represents 10 μm. **b**, Optical micrograph of sample **K2**. The scale bar represents 5 μm. **c**, Representative Raman spectra of **K2** obtained before (black line) and after (red line) thermal annealing at 400 °C for 2 hours. The absence of the disorder-related D band (marked by the vertical arrow) indicates high crystalline order of the sample. **d,** Raman map for the G mode frequency ($\omega_G$) obtained from pristine **K1**. **e,** Raman map for the G mode frequency ($\omega_G$) obtained from pristine **K2**. **f,** Raman map for the G mode frequency ($\omega_G$) obtained from annealed **K1**. **g,** Raman map for the G mode frequency ($\omega_G$) obtained from annealed **K2**. Each of the Raman mapping was carried out in the areas specified by the blue dashed boxes spanning 20x20 and 35x15 μm$^2$ for **K1** and **K2**, respectively. Each Raman spectrum of the map data was obtained for 15 s.

**Figure 2. Correlation between the frequencies of the G and 2D Raman modes of graphene ($\omega_G$, $\omega_{2D}$).** The data were obtained from Raman mapping of three graphene samples (**K1**, **K2** & **K3**, respectively, in red, blue & black) before (+) and after (×) thermal annealing at 400 °C. Each Raman spectrum of the map data was obtained for 15 s. The green dot (denoted **O**) obtained from a freestanding graphene sample (**F1**) represents ($\omega_G^0$, $\omega_{2D}^0$) which is not affected by strain or charge doping. (See the text and also Supplementary Figure S1.) The red and blue solid lines represent ($\omega_G$, $\omega_{2D}$) of graphene doped with varying density of holes and electrons, respectively, induced by an electrical method (Ref. 49). (It is to be noted that there is an equivalent work, Ref. 43, which shows some discrepancy from Ref. 49. For detailed discussion, see Supplementary Methods A.) The magenta dashed line is an average of experimental ($\omega_G$, $\omega_{2D}$) for strain-free graphene with varying density of holes (*n*) (Refs. 43 & 49). The black dashed line represents a prediction of ($\omega_G$, $\omega_{2D}$) for charge-neutral graphene under randomly oriented uniaxial stress. (See the text.) **Inset,** Decomposition of the effects of hole doping and strain using a vector model. Any given ($\omega_G$, $\omega_{2D}$), **OP**, can be decomposed into **OH** along the "strain-free" unit vector, $e_H$ for hole doping, and **OT** along the "charge-neutral" unit vector, $e_T$ for tensile strain (-$e_T$ for compressive strain), respectively. $e_H$ and $e_T$ divide the $\omega_G$-$\omega_{2D}$ space into the four quadrants ($Q_1 \sim Q_4$).

**Figure 3. Refined analysis model considering the logarithmic phonon anomaly. a**, The effects of hole carriers with varying density *n* on ($\omega_G$, $\omega_{2D}$) of graphene. The red solid line presents the experimental trajectory of ($\omega_G$, $\omega_{2D}$) as a function of *n* obtained by Das. et al. (Ref. 49). The red vertical bars on the red solid line mark *n*, every $1.0 \times 10^{12}$ cm$^{-2}$. The blue solid line represents a theoretical prediction by Lazzeri et al. reported in Ref. 35 which considered the effects of the logarithmic phonon anomaly. We used the relationship that $\omega_G(n) = \omega_G^o + \Delta\omega_G(n)$ and $\omega_{2D}(n) = \omega_{2D}^o + \Delta\omega_{2D}(n)$, where ($\omega_G^o$, $\omega_{2D}^o$) corresponds to **O** of Fig. 2. The numerical values of $\Delta\omega_{2D}(n)$ were obtained assuming its linear relationship with $\Delta\omega_G(n)$ as shown in Supplementary Figure S3b. The hole density (*n*) for the blue solid line ranges from 0 (corresponding to **O**) to ~$2.6 \times 10^{12}$ cm$^{-2}$ as marked by the yellow and magenta diamonds every $0.2 \times 10^{12}$ and $1.0 \times 10^{12}$ cm$^{-2}$, respectively. **b**, The theoretical and experimental trajectories of ($\omega_G$, $\omega_{2D}$) affected by *n* ranging from 0 to $1.6 \times 10^{13}$ cm$^{-2}$ with Raman map data of pristine (+) and annealed (×) **K2**. The red solid line is identical to the one in

Fig. 3a. The black horizontal bars on the $e_T$ line designate uniaxial strain ($\varepsilon$) ranging from -0.6 to 0.3%, each bar representing a step of 0.1%. The uniaxial strain-sensitivity of the G mode, $\Delta\omega_G/\Delta\varepsilon$ = -23.5 cm$^{-1}$/%, was calculated from the work by Yoon et al. (Ref. 29) considering the splitting of the G mode and random crystallographic orientation of strain. In case of biaxial strain, however, an averaged sensitivity factor of -69.1 ± 3.4 cm$^{-1}$/% can be used (Ref. 27, 30 & 47).

**Figure 4. Thermally induced variations in ($\omega_G$, $\omega_{2D}$) upon successive annealing.** The data were obtained from Raman mapping of a graphene sample (**K4**) before and after successive thermal annealing at various temperatures ($T_{anneal}$ in $^o$C), except those in dark yellow which were obtained from **K3** following the one-time annealing at 400 $^o$C (see Fig. 2). The horizontal and vertical error bars for $T_{anneal}$ = 400 & 500 $^o$C represent standard deviations in $\omega_G$ and $\omega_{2D}$, respectively. Each Raman spectrum of the map data was obtained for 10 s.

**Figure 5. Decomposition of the effects of strain concurrent with spatially varying charge doping in pristine graphene. a,** Optical micrograph of a graphene sample (**K5**) under tensile stress concurrent with charge doping. The scale bar represents 10 μm. **b,** Raman map of $\omega_G$. **c,** Raman map of $\omega_{2D}$. **d,** Raman map for $\Gamma_G$. **e,** Raman map of $\omega_{G,\,\varepsilon=0}$. **f,** Raman map of $\omega_{G,\,n=0}$. **g,** Correlation between $\omega_G$ and $\omega_{2D}$ which were given in **b** & **c**. The Raman map data in **a** ~ **c** were obtained from pristine **K5** for 10 s/pixel from the area (20x15 μm$^2$) indicated by the blue dashed box in **a**. The data in **e** & **f** were mathematically derived by the simple vector model described in Fig. 2. (See the text.)

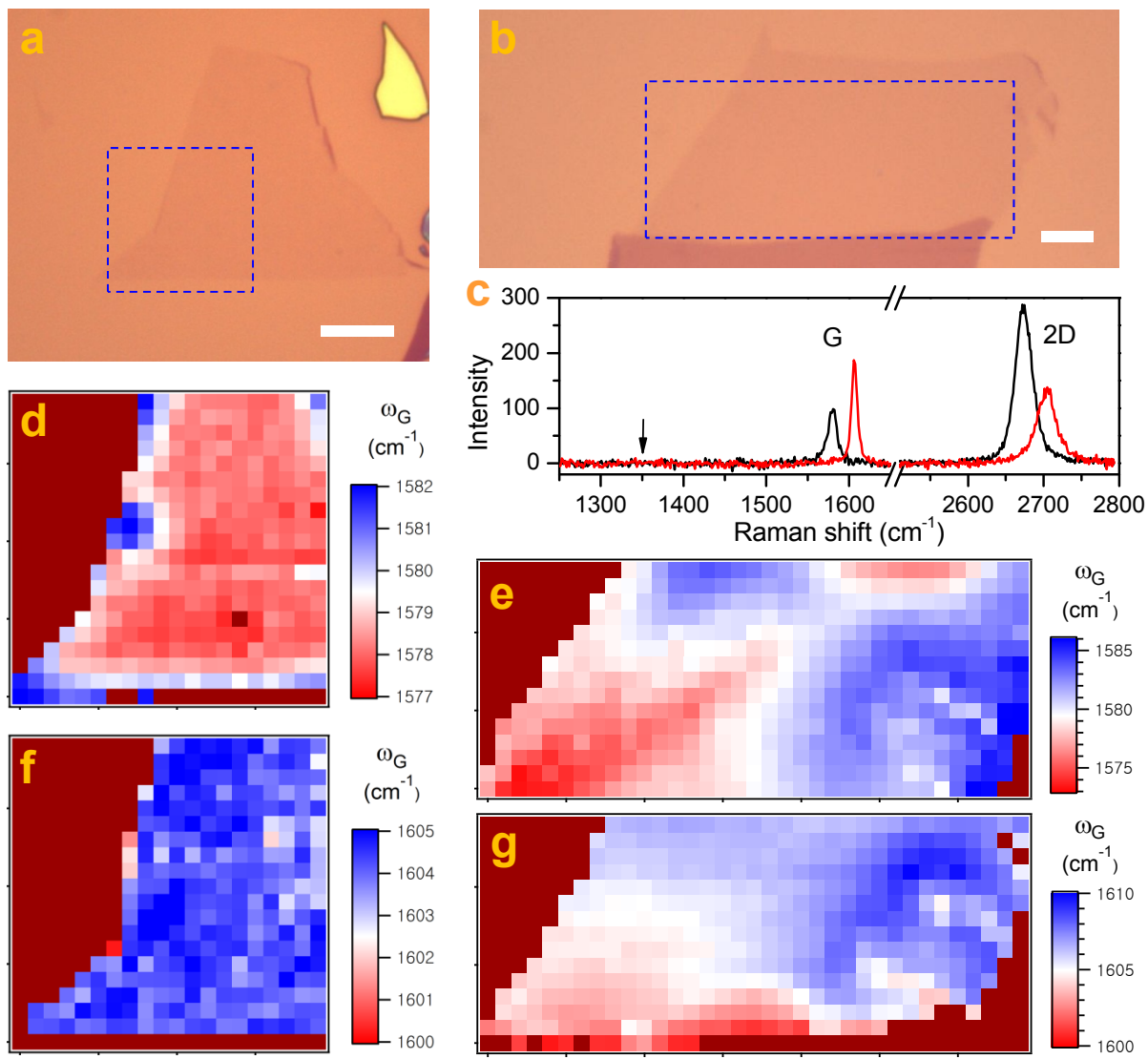

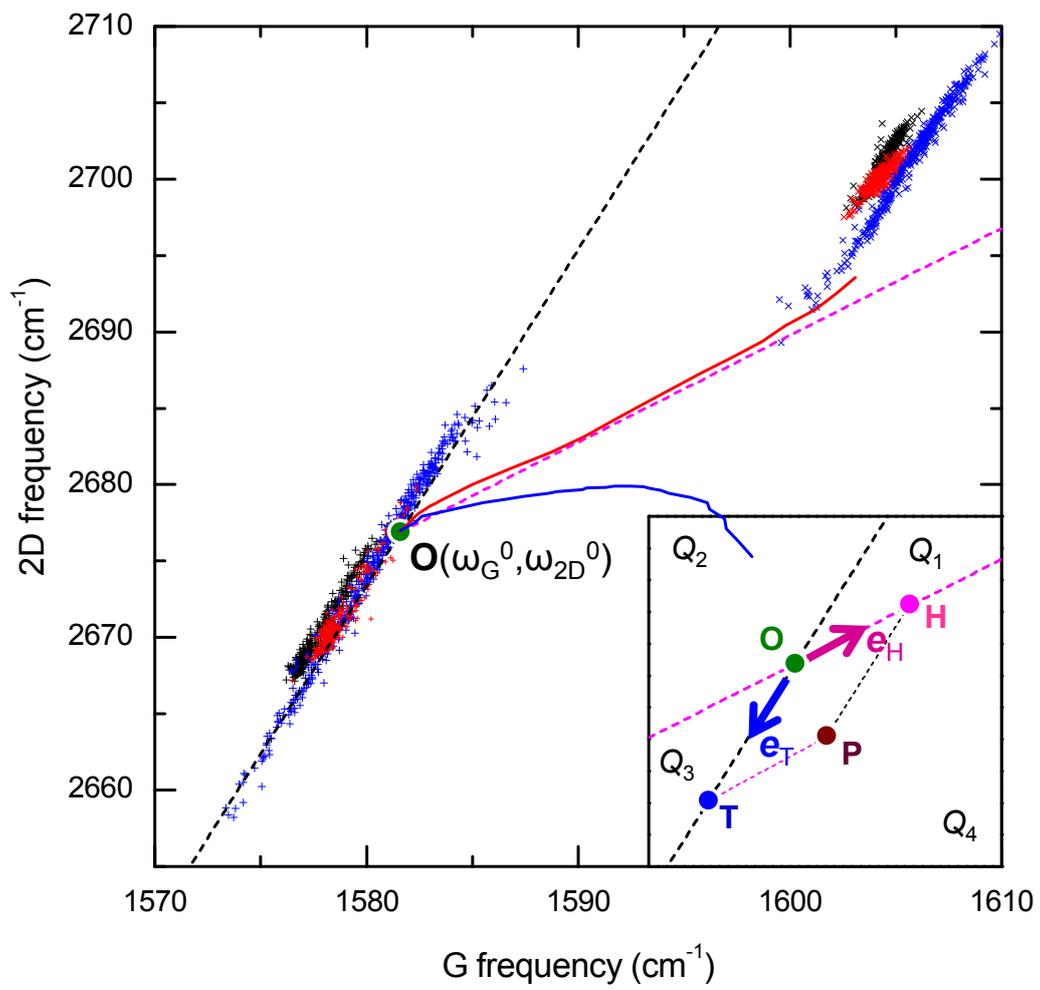

**Figure-2 (Ryu)**

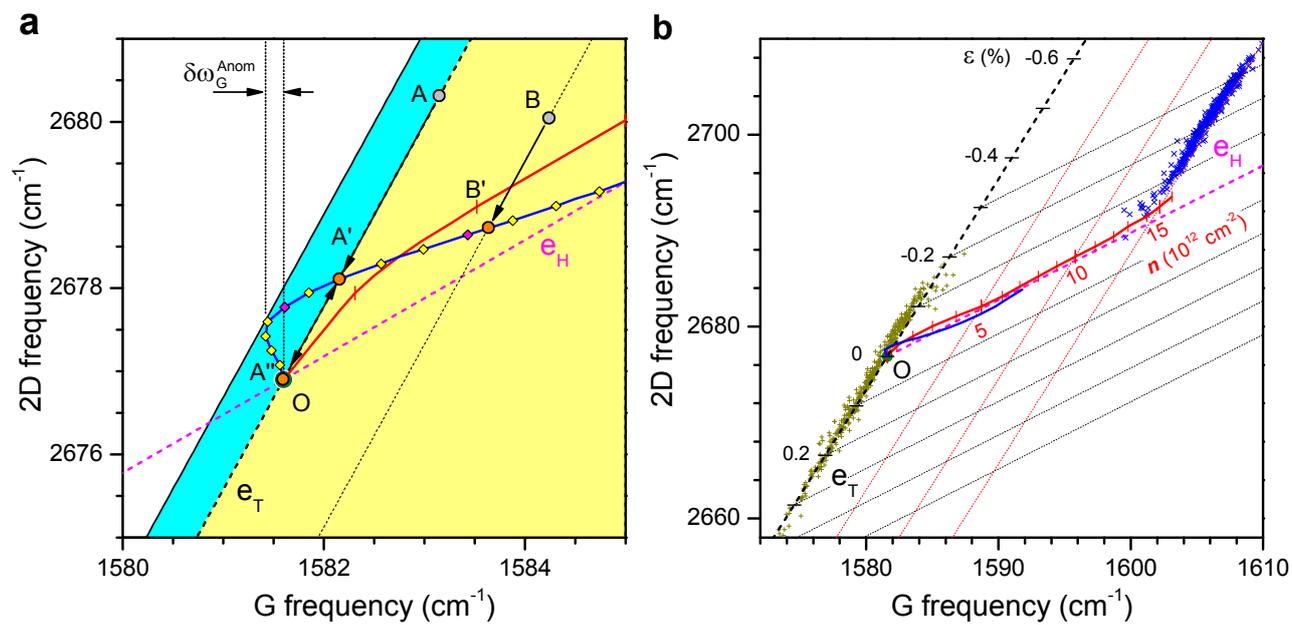

**Figure-3 (Ryu)**

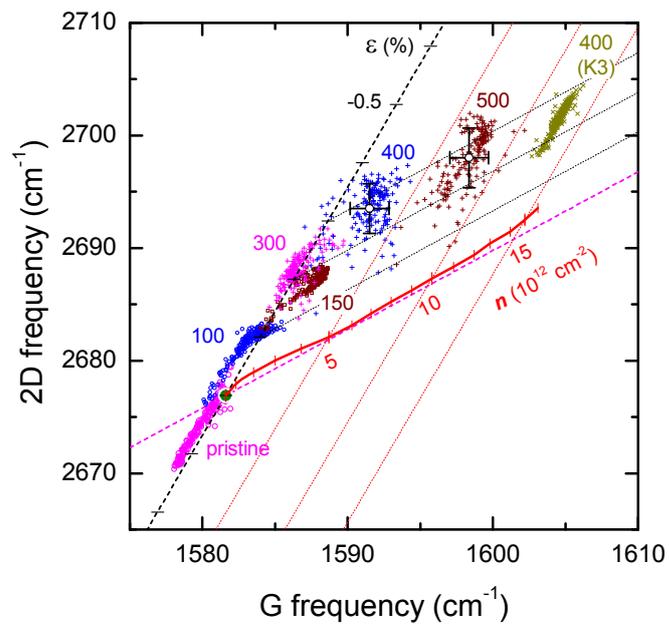

**Figure-4 (Ryu)**

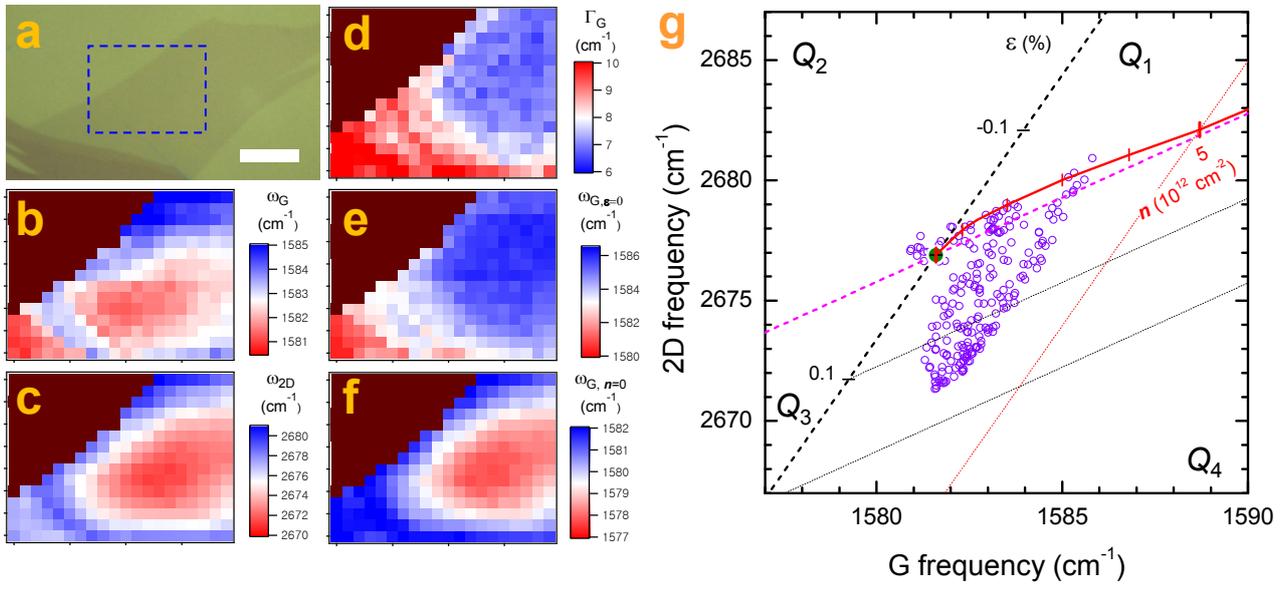

**Figure-5 (Ryu)**

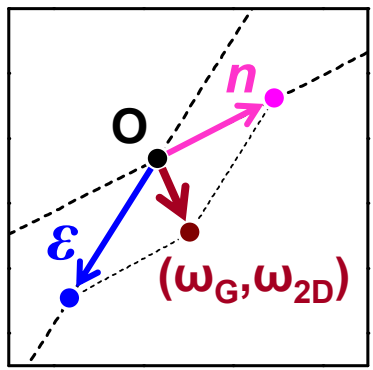

**Thumbnail (Ryu)**